\documentclass[conference]{IEEEtran}
\ifCLASSINFOpdf
   \usepackage[pdftex]{graphicx}

   \graphicspath{{../pdf/}{figures/}}
   \DeclareGraphicsExtensions{.pdf,.jpeg,.png,.jpg}
\else
\fi
\usepackage{amssymb}
\usepackage[cmex10]{amsmath}
\usepackage{algorithmic}
\usepackage{algorithm}
\usepackage{color}
\usepackage[table]{xcolor}
\usepackage{multicol}
\usepackage{multirow}

\usepackage{mdwmath}
\usepackage{mdwtab}
\usepackage{url}

\usepackage{array}
\newcolumntype{L}[1]{>{\raggedright\let\newline\\\arraybackslash\hspace{0pt}}m{#1}}
\newcolumntype{C}[1]{>{\centering\let\newline\\\arraybackslash\hspace{0pt}}m{#1}}
\newcolumntype{R}[1]{>{\raggedleft\let\newline\\\arraybackslash\hspace{0pt}}m{#1}}

\usepackage[tight,footnotesize]{subfigure}
\def\Min{{\rm minimize }}

\def\b0{{\bf 0}}
\def\bb1{{\bf 1}}

\usepackage[printonlyused]{acronym}

\acrodef{MAS}{Multi-Agent System}
\acrodef{DC}{Data Center}
\acrodef{VRAN-PAP}{VRAN Placement and Assignment Problem}
\acrodef{FPP}{Function Placement Problem}
\acrodef{SPP}{Server Placement Problem}
\acrodef{VDCE}{Virtual Data Center Embedding}
\acrodef{ILP}{Integer Linear Program}
\acrodef{BILP}{Binary Integer Linear Program}
\acrodef{MILP}{Mixed Integer Linear Program}
\acrodef{LARE}{Langragian Relaxation}
\acrodef{VNE}{Virtual Network Embedding}
\acrodef{OTT}{over-the-top}
\acrodef{BPP}{Bin Packing Problem}
\acrodef{MINLP}{Mixed-Integer Nonlinear Programming}
\acrodef{ML}{Machine Learning}
\acrodef{LP}{Linear Program}
\acrodef{CFLP}{Capacitated Facility Location Problem}
\acrodef{ISG}{Industry Standards Group}
\acrodef{ETSI}{European Telecommunications Standards Institute}
\acrodef{PM}{Physical Machine}
\acrodef{VM}{Virtual Machine}
\acrodef{EPC}{Evolved Packet Core}
\acrodef{MCLP}{Maximal Covering Location Problem}
\acrodef{CPRI}{Common Public Radio Interface}
\acrodef{VRAN}{Virtualized Radio Access Network}
\acrodef{LPTRA}{Low Power Transmit and Receive Antenna}
\acrodef{HPTRA}{High Power Transmit and Receive Antenna}
\acrodef{NFV-HRAN}{NFV-based HRAN}
\acrodef{RRH}{Remote Radio Head}
\acrodef{BBP}{Base Band Processing}
\acrodef{BBU}{Base Band Unit}
\acrodef{vBBU}{Virtualized Base Band Unit}
\acrodef{eNodeB}{Evolved Node B}
\acrodef{UE}{User Equipment}
\acrodef{PN}{Physical Network}
\acrodef{VN}{Virtual Network}
\acrodef{HRAN}{Heterogeneous Radio Access Network}
\acrodef{RAN}{Radio Access Network}
\acrodef{CRAN}{Centralized Radio Access Network}
\acrodef{TSP}{Telecommunication Service Provider}
\acrodef{BS}{Base Station}
\acrodef{CSI}{Channel State Information}
\acrodef{QSI}{Queue State Information}
\acrodef{NFV}{Network Function Virtualization}
\acrodef{HPN}{High Power Node}
\acrodef{LPN}{Low Power Node}
\acrodef{RL}{Reinforcement Learning}
\acrodef{LTE}{Long Term Evolution}
\acrodefplural{CAPEX}[CAPEX]{CAPital EXpenses}
\acrodefplural{OPEX}[OPEX]{OPerating EXpenses}

\begin{document}
%
\title{Placement and Assignment of Servers in Virtualized Radio Access Networks}

\author{\IEEEauthorblockN{Rashid Mijumbi\IEEEauthorrefmark{1},
Joan Serrat\IEEEauthorrefmark{1},
Juan-Luis Gorricho\IEEEauthorrefmark{1},
Javier Rubio-Loyola\IEEEauthorrefmark{3} and
Steven Davy\IEEEauthorrefmark{2}
}
\IEEEauthorblockA{\IEEEauthorrefmark{1}Universitat Polit\`{e}cnica de Catalunya, 08034 Barcelona, Spain}
\IEEEauthorblockA{\IEEEauthorrefmark{3}CINVESTAV, Tamaulipas, Mexico}
\IEEEauthorblockA{\IEEEauthorrefmark{2}Telecommunications Software and Systems Group, Waterford Institute of Technology, Ireland}
}

\maketitle

\begin{abstract}
The virtualization of \acp{RAN} has been proposed as one of the important use cases of \ac{NFV}. In \acp{VRAN}, some functions from a \ac{BS}, such as those which make up the \ac{BBU}, may be implemented in a shared infrastructure located at either a data center or distributed in network nodes. For the latter option, one challenge is in deciding which subset of the available network nodes can be used to host the physical \ac{BBU} servers (the placement problem), and then to which of the available physical \acp{BBU} each \ac{RRH} should be assigned (the assignment problem). These two problems constitute what we refer to as the \ac{VRAN-PAP}. In this paper, we start by formally defining the \ac{VRAN-PAP} before formulating it as a \ac{BILP} whose objective is to minimize the server and front haul link setup costs as well as the latency between each \ac{RRH} and its assigned \ac{BBU}. Since the \ac{BILP} could become computationally intractable, we also propose a greedy approximation for larger instances of the \ac{VRAN-PAP}. We perform simulations to compare both algorithms in terms of solution quality as well as computation time under varying network sizes and setup budgets.
\end{abstract}


\section{Introduction}

\begin{figure}[t]
\centering
\includegraphics[width=9.5cm]{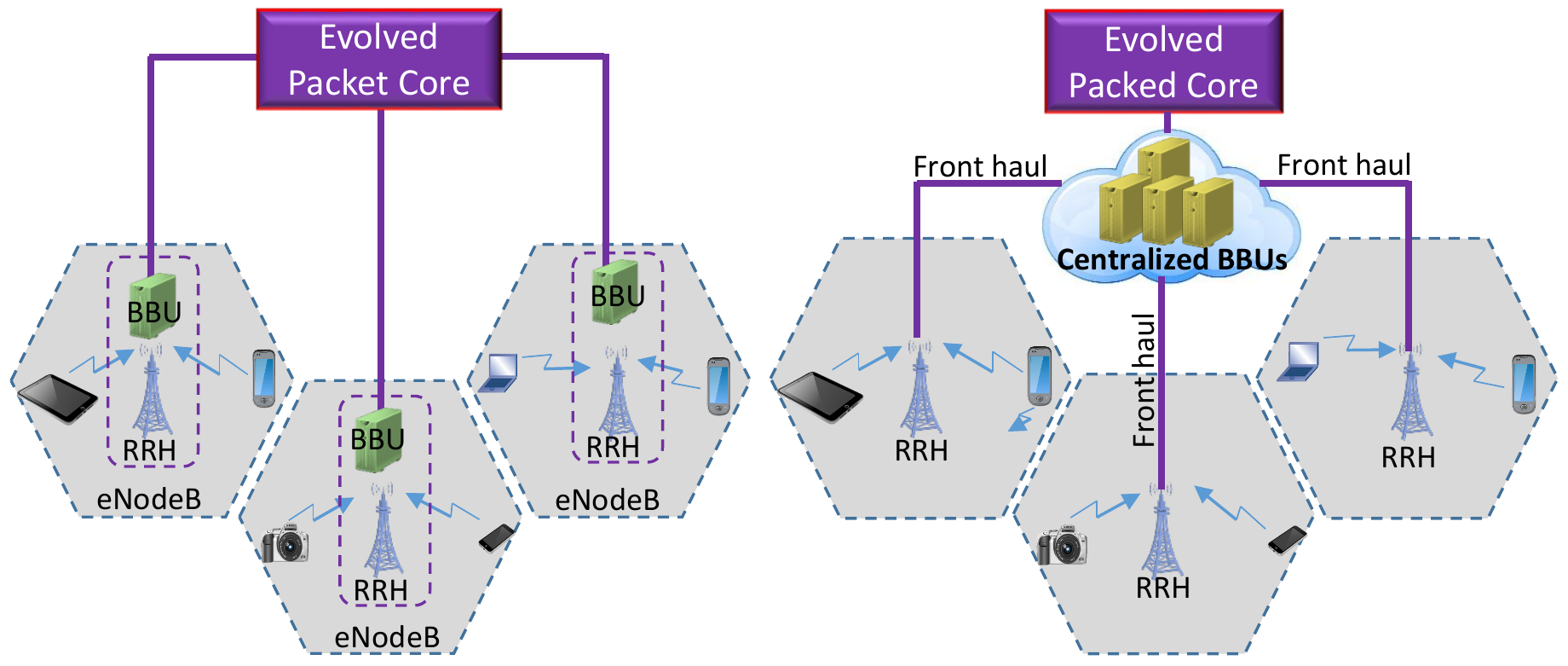}
  \caption{Virtualization of LTE Radio Access Networks}
  \label{current}
\end{figure}

The telecommunications sector continues to be faced with an apparent insatiable demand for higher data rates by subscribers. This explosive traffic demand is forecast to continue doubling every year to 2020 \cite{nfv}. To keep up with this demand, new infrastructure is often needed, which leads to high \acp{CAPEX} and \acp{OPEX} for \acp{TSP}. However, due to competition both against each other and from services provided \ac{OTT} on their data channels, \acp{TSP} cannot respond to the increases in \acp{CAPEX} and \acp{OPEX} with increased subscription fees. This has significantly reduced profitability, and forced \acp{TSP} to find new ways of expanding the capacity of their networks while still remaining profitable \cite{MijumbiNFV15}. These new solutions need to identify parts of the network which contribute highly to \acp{CAPEX} and \acp{OPEX}, and to find ways of efficiently utilizing their resources.\\
\indent \acp{eNodeB} are usually dimensioned so as to be able to handle peak-hour traffic, leading to inefficient resource utilization during non-peak hours, or due to mobility of subscribers \cite{ChinaMobile}. In addition, up to 80\% of \acp{CAPEX} and 60\% of \acp{OPEX} is spent on \acp{RAN} \cite{ChinaMobile}. These factors have made \acp{RAN} an important candidate for achieving reduced \acp{CAPEX} and \acp{OPEX} through more efficient resource utilization. This can be achieved by employing virtualization technology and network programming to decouple \ac{RAN} functions from the physical devices on which they run. This is the concept of \ac{nfv, mano} \cite{MijumbiNFV15}, which continues to draw immense attention from not only standardization bodies but also researchers in both industry and academia.\\
\indent In the proposed \ac{VRAN} \cite{ETSIUseCases}, \ac{RAN} functions such as the \ac{BBU} $-$ including \ac{BBP}, Media Access Control (MAC), Radio Link Control (RLC), Packet Data Convergence Protocol (PDCP), Radio Resource Control (RRC), Control and Coordinated Multi-Point (CoMP) transmission and reception $-$ can be dispatched to a \ac{TSP} as an instance of plain software. The left part of Fig. \ref{current} shows the current implementation of a \ac{RAN}. It can be observed that in a traditional implementation, the \ac{RRH} in each cell is associated with a dedicated \ac{BBU}, which is then connected to the \ac{EPC}. In the \ac{VRAN} scenario (the right part of Fig. \ref{current}), the servers responsible for the \ac{BBU} functions are transferred to a shared physical infrastructure and virtualized. They are then connected to the \acp{RRH} over front haul links $-$ possibly based on optical fiber. The physical \ac{BBU} servers may be located at a data center or distributed across the \ac{TSP}'s \ac{RAN} node locations.\\
\indent However, despite the promising gains, \acp{VRAN} present some challenges especially with regard to front haul links, which are not only capital intensive, but also introduce high latency \cite{ChihLin14}. In order to achieve the gains expected from \acp{VRAN}, i.e. energy and spectral efficiency, front haul links must be able to provide a high bandwidth, at low capital and operating costs and low latency. To this end, there have been efforts to enhance the efficiency of sharing the constrained and expensive front haul resources \cite{ChihLin14, SeokPark13, LiJPeng14}. However, to the best of our knowledge, all current approaches assume that the physical BBU servers have already been setup, and \ac{RRH} sites assigned to them. The total number of physical \ac{BBU} servers required to serve the whole \ac{RAN}, together with their location relative to the \ac{RRH} sites they serve is important since it does not only determine costs of server and front haul link deployment and maintenance, but also the resulting latency between each \ac{RRH} and its assigned \ac{BBU} server. Therefore, the design of efficient placement and assignment algorithms is an important initial step towards achieving the expected \acp{CAPEX} and \acp{OPEX} gains, as well ensuring that the resulting latency is acceptable.\\
\indent In this paper, we propose algorithms for placing \ac{BBU} servers and assigning \ac{RRH} sites to them. In particular, we consider a \ac{TSP} that already has an existing topology of \acp{BS}. To virtualize the RAN, the TSP should determine a subset of its existing network nodes where physical \acp{BBU} may be placed. In addition, for each of the \acp{RRH} at sites where no physical BBU has been placed, we should assign them to one of the placed \acp{BBU}. These two problems constitute what we refer to as the \ac{VRAN-PAP}. The placement and assignment should be performed based on a set of constraints. For example, there is a maximum budget which restricts the number of physical BBUs which can be placed. In addition, the assignment should ensure that physical server capacity is not exceeded and that front haul link latency is minimized.\\
\indent The main contributions of this paper are three-fold: (1) To the best of our knowledge, this is the first attempt to define the \ac{VRAN-PAP}, (2) a \ac{BILP} formulation of the \ac{VRAN-PAP} which may be used for smaller problem instances or where high accuracy is required, and (3) a greedy approximation which may be used to solve larger instances of the \ac{VRAN-PAP}. Being the first foray into defining, formulating and solving of the \ac{VRAN-PAP}, we hope that the proposed algorithms may be used as benchmarks for future work in the same area.\\
\indent The rest of the paper is organized as follows: We discuss related work in section \ref{related} before describing the \ac{VRAN-PAP} in section \ref{desc}. In section \ref{ilp}, we formulate the \ac{VRAN-PAP} as a \ac{BILP} and propose a greedy approximation for it in section \ref{greedy}. The two algorithms are evaluated and discussed in Section \ref{eval}, and the paper concluded in Section \ref{conc}.


\section{Related Work} \label{related}
To be able to capture both the maximum number of BBU servers, cost and latency objectives, the \ac{VRAN-PAP} defined in this paper involves a non-trivial integration of ideas from both the \ac{MCLP} \cite{MCLP74} as well as the \ac{CFLP} \cite{vygen2005approximation}. It may also be reduced to the classical \ac{BPP} \cite{Kenyon97}. However, in \ac{VRAN}, front haul links will have costs that must be considered. In addition, compared to each of the above problems, the \ac{VRAN-PAP} presents more design considerations such as cost, scalability, latency and resource utilization efficiency. These considerations do not only provide an extra degree of freedom which may be used to align the solution towards specific NFV objectives (such as dynamic resource scaling), but also increase the complexity of the problem.\\
\indent The management of resources in \acp{CRAN} has been considered in \cite{VuNguyen14, JianHua14}, where the problem is formulated as an optimization problem with varying objectives and solved using different approaches. Yegui et al. \cite{YeguiCia14} apply a decision-theoretic framework to the topology configuration and rate allocation problem in \ac{CRAN} with the objective of optimizing the end-to-end performance of the users, while Yuanming et al. \cite{Yuanming14} propose a framework to design a ``green Cloud-RAN". They formulate the problem as a joint \ac{RRH} selection and power minimization beamforming problem, and solve it using both greedy and group sparse beamforming algorithms. FluidNet \cite{Sundaresan135} is a framework for dynamically re-configuring the front haul of a \ac{CRAN} to meet the dual objective of improved \ac{RAN} performance with reduced resource usage in the BBU pool. While this is not an exhaustive list of all proposals with regard to resource management in \acp{CRAN}, we note that unlike our proposal these approaches are mainly focused on either front haul or back haul resource allocation and not the placement of servers that creates these resources in the first place.\\
\indent The placement of servers in \acp{VRAN} is related to the \ac{FPP} \cite{MijumbiNFV15}. The \ac{FPP} involves the need to determine on to which physical resources (servers) network functions are \textit{placed}, and be able to move functions from one server to another for such objectives as load balancing, energy saving, recovery from failures, etc. The \ac{FPP} can be formulated as an optimization problem, with a particular objective. Such an approach has been followed by \cite{BastaA2014, Moens14, Bagaa14}. Since optimal solutions could become intractable, some authors \cite{MijumbiNFV15, Xia15, Yoshida14} have proposed heuristics for the same purpose. Similarly, Rappaport et al. \cite{Rappaport13} formulate an optimization problem based on the soft-\ac{CFLP} with the objective of placing applications and their related data over a cloud infrastructure. Finally, the placement of servers is related to \ac{VDCE} \cite{BariCST13} and \ac{VNE} \cite{ShidIM15, path, rl, aims, sdn, neurofuzzy, neural}, both of which are well studied problems. With regard to \ac{VDCE}, most current approaches such as \cite{MengXia10, RabbaniMG13} focus on resource sharing through mapping \acp{VM} to physical servers with the aim of improving server resource (e.g., CPU, memory and disk) utilization, and maximizing the number of mapped \acp{VM}. Similarly, \ac{VNE} deals with mapping nodes and links of a \ac{VN} onto nodes and and links in a \ac{PN}, with the objective of embedding as many \acp{VN} as possible onto a given substrate.

\section{Problem Description} \label{desc}
\begin{figure}[t]
\centering
{\includegraphics[width=8.5cm, height=6.0cm]{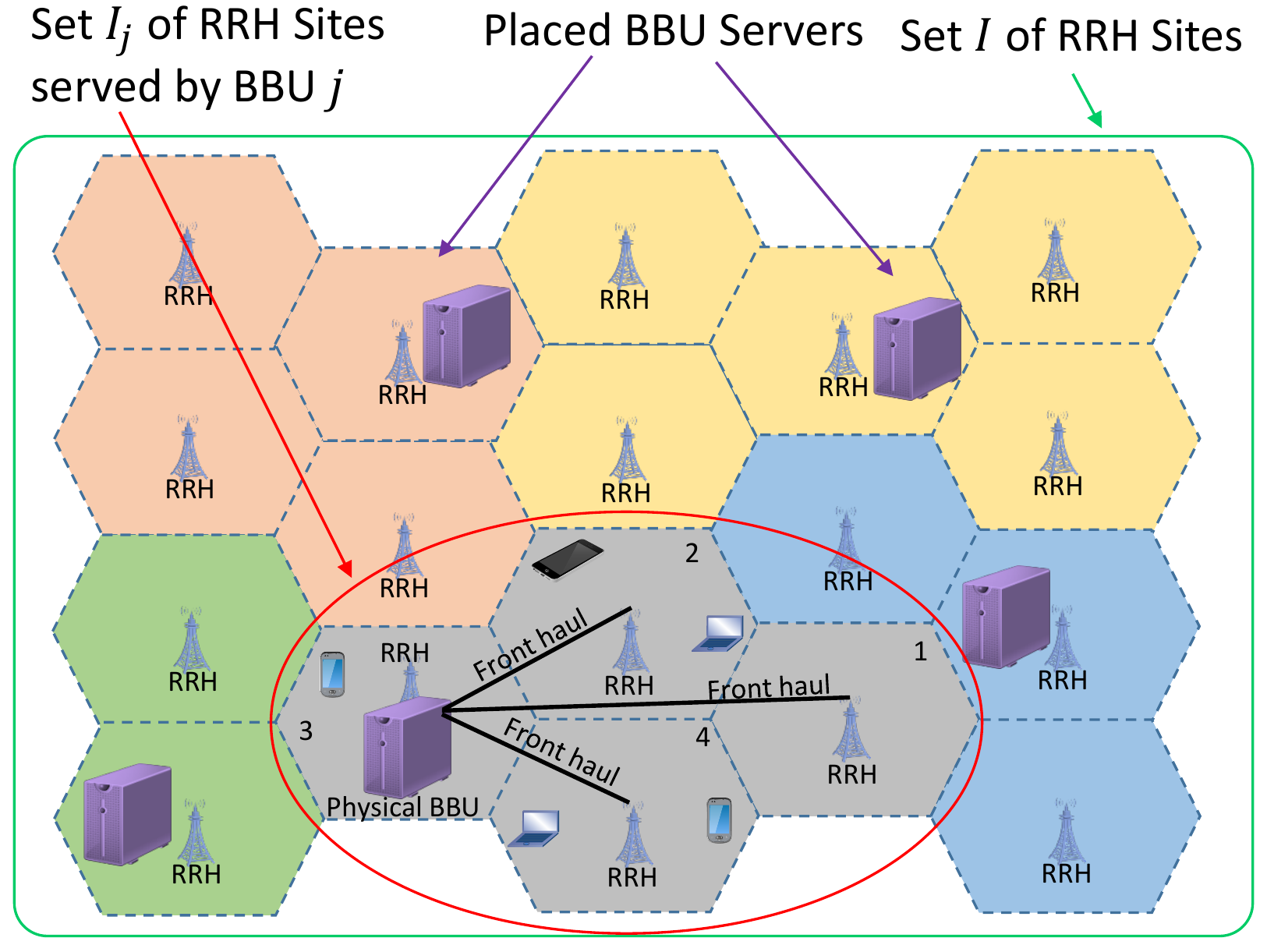}}
  \caption{Placement and Assignment Problem Representation}
  \label{problem}
\end{figure}

The \ac{VRAN-PAP} considered in this paper is represented in Fig. \ref{problem}. In the figure, a number of \ac{RRH} sites  are grouped together, as represented by cells with the same color. All RRH sites in a given group are \textit{assigned} to a physical BBU server \textit{placed} in one of the cells in that group. As an example, it can be seen from the figure that cells 1, 2, 3 and 4 share a common BBU server located in cell 3. For this group of RRHs, there are front haul links from cells 1, 2 and 4 to cell 3. In what follows, we propose a representation of the three main components of the scenario shown in Fig. \ref{problem}.
\subsection{\ac{RRH} Sites}
We consider a \ac{RAN} with a set $I$ of \ac{RRH} sites that must be served by another set $J$ of BBU servers, where $|J| \leq |I|$. The location of each \ac{RRH} site $i \in I$ is given by $P_i(x,y)$, where $x$ may be the latitude, and $y$ the longitude. Each \ac{RRH} site $i \in I$ provides service to $U_i$ associated \acp{UE} whose combined average processing requirements are $\delta_i$. For each \ac{RRH} site, we define a maximum desired level of latency $\tau_i$, which gives a measure of the desired maximum distance from the \ac{RRH} to the \ac{BBU}. 
\subsection{\ac{BBU} Servers}
Due to budget restrictions, we consider that a TSP is able to set up a maximum of $p = |J|\leq |I|$ physical BBU servers to serve all the RRH sites in the RAN. Each physical server $j \in J$ has a maximum processing capacity of $\eta_j$. This maximum processing capacity is aimed at reflecting the fact that in practice, we would not be able to have a physical \ac{BBU} server (or multiples of them in a single location) with unlimited processing capacity. Placing a physical BBU $j \in J$ involves a cost of $f_j$, which may be defined as a linear-cost function shown in \eqref{costfn}. The cost is composed of two parts: a fixed initial cost $u_j$ which takes care of the fixed investments such as space and installations, and a marginal/incremental cost $v_j$ per unit of the processing capacity installed at the BBU server.

\begin{equation}
f_j = u_j + v_j \eta_j
\label{costfn}
\end{equation}
Each \ac{BBU} server is located at the site of one of the \acp{RRH}. Each subset $I_j \subseteq I$ of RRH sites is served by a single BBU server $j$ located in one of the \ac{RRH} sites.
\subsection{Front Haul Links}
Each \ac{RRH} site $i$ is connected to the corresponding \ac{BBU} server $j$ by a front haul link $l_{ij} \in L$, where $L$ is the set of all front haul links.
Each front haul link $l_{ij}\in L$ has a latency $t_{ij}$, which is dependent on the distance $d_{ij}$ between the site $i$ and the server $j$ and the speed of signals in the transport medium used. The cost of setting up a front haul link between a site $i$ and a physical server $j$ is $c_{ij}$, and is defined as a linear combination of an initial fixed cost $\omega_{ij}$ and a variable part dependent on the bandwidth $B_{ij}$ required on the link as shown in \eqref{costdef}. In turn, we define the bandwidth $B_{ij}$ required for a front haul link between \ac{BBU} $j$ and \ac{RRH} $i$ as being directly related to the processing requirements $\delta_i$ of the \ac{RRH} site, and is hence given by $B_{ij} = \gamma \delta_i$, where $\gamma$ and $\chi$ are a constants.
\begin{equation}
c_{ij} = \omega_{ij} + \chi B_{ij}
\label{costdef}
\end{equation}
For a given solution, the maximum latency that any RRH site would experience while communicating with its assigned physical \ac{BBU} is a measure of the worst possible latency performance of the system.

\begin{table}[t]
\caption{Summary of Problem Parameters and Decision Variables}
\label{params}
\renewcommand{\arraystretch}{1.5}
\centering
\rowcolors{2}{gray!25}{white}
\begin{tabular}{ C{1.6cm} L{6.5cm} } \hline
\bfseries  Symbol & \bfseries Description\\ \hline \hline
$i \in I $ & Index for the set of \acp{RRH} sites that must be served\\
$j \in J$ & Index for the set of potential \ac{BBU} servers $(|J| \leq |I|)$\\
$p = |J|$ & Maximum number of physical \ac{BBU} servers that may be installed\\
$\delta_i > 0$ & Average processing requirements of all subscribers using the \ac{RRH} at site $i$ \\
$\eta_j > 0 $ & Processing capacity of a physical server placed at location $j$ \\
$\tau_i > 0$ & Desired maximum latency for users using \ac{RRH} site $i$\\
$f_j \geq 0 \in \mathbb{R}_+ $ & Cost of placing and/or installing a physical server at location $j$ \\
$c_{ij} \geq 0 \in \mathbb{R}_+ $ & Cost of establishing the front haul link $l_{ij}\in L$ between \ac{RRH} $i$ and \ac{BBU} $j$\\
$t_{ij} > 0$ & Actual latency for a front haul link $l_{ij}\in L$ connecting \ac{RRH} $i$ and \ac{BBU} $j$\\
$y_{j}$ & Binary variable that takes on value $1$ if \ac{BBU} $i$ exists, $0$ otherwise \\
$x_{ij}$ & Binary variable that takes on value $1$ if \ac{RRH} $i$ is assigned to \ac{BBU} $j$, 0 otherwise \\

\hline
\end{tabular}
\end{table}

\section{\ac{BILP} Formulation} \label{ilp}
\subsection{Definition of Decision Variables}
Let $y_j$ be a binary decision variable that takes on a value of 1 if a physical BBU server $j$ should be setup, and $0$ otherwise. In addition, we define $x_{ij}$ as a binary variable that takes on value $1$ if the RRH site $i \in I$ is assigned to a BBU server $j \in J$. The task is to assign values to all possible occurrences of $y_j$ and $x_{ij}$, such that both latency and total (server + front haul link) costs are minimized. For ease of reference, we summarize all parameters and decision variables in Table \ref{params}.

\subsection{Objective}

\begin{align}
\label{obj1}
\begin{split}
\Min \hspace{5 mm} \alpha \Bigg( \sum_{j \in J} \Big( f_j \times y_j \Big) + \sum_{i \in I} \sum_{j \in J} \Big( c_{ij} \times x_{ij} \Big) \Bigg) + \\ \beta \sum_{i \in I} \sum_{j \in J} \Big(t_{ij} - \tau_i \Big) x_{ij}
\end{split}
\end{align}

The objective in \eqref{obj1} is to minimize the latency as well as costs. In particular, the first term of the objective gives the total costs of installing the \ac{BBU} servers, while the second term includes the costs of setting up front haul links between each \acp{RRH} and the \ac{BBU} that serves it. Finally, the last term gives the total deviation of the actual latency from the desired levels between each \ac{RRH} and the corresponding \ac{RRH}. While both latency and front haul link cost depend on the length of the link, the motivation of the third term in the objective is that irrespective of cost, it is possible to use it to give better priority QoS to users served by certain \ac{RRH} sites by defining their desired latencies as being very low, and hence ensuring that they get served by the closest \acp{BBU}. The constants $\alpha$ and $\beta$ may be used not only to give more importance either to costs or to latency, but also to scale the values of costs to be comparable to those of latency so as to have a meaningful summation.

\subsection{Constraints}

\subsubsection{Placement and Assignment}

\begin{equation}
\sum \limits_{j \in J} x_{ij} = 1 \hspace{10 mm}\forall i \in I
\label{c2}
\end{equation}

\begin{equation}
x_{ij} - y_i \leq 0  \hspace{10 mm}\forall i \in I, j \in J
\label{c3}
\end{equation}

Constraint \eqref{c2} ensures that each \ac{RRH} is attached to one physical BBU server. This constraint ensures that each \ac{RRH} is assigned to at least one eligible \ac{BBU} by creating a front haul link between them. Constraint \eqref{c3} ensures that a front haul link is created between a RRH site $i$ and a \ac{BBU} server $j$, only if $j$ has been placed. Together, constraints \eqref{c2} and \eqref{c3} ensure that the required number of \acp{BBU} are placed to serve all RRH sites, and that \acp{RRH} are only assigned to sites where \acp{BBU} have been placed. 

\subsubsection{Resource Capacity and Budget Conservation}

\begin{equation}
\sum \limits_{j \in J} y_{j} \leq p
\label{c1}
\end{equation}

\begin{equation}
\sum \limits_{i \in I}  \Big(\delta_i \times x_{ij} \Big) \leq \eta_j \hspace{10 mm}\forall j \in J
\label{c4}
\end{equation}

Constraint \eqref{c1} ensures that the maximum number of \ac{BBU} servers does not exceed the budget $p$, while \eqref{c4} is a server capacity constraint that ensures that the total processing requirements of all \acp{RRH} assigned to a given \ac{BBU} server do not exceed the actual physical resources installed at the server. Finally, \eqref{c5} and \eqref{c6} give the binary domain variables.

\subsubsection{Domain Constraints}

\begin{equation}
x_{ij} \in \big\{ 0 , 1 \big\} \hspace{10 mm}\forall j \in J, i \in I
\label{c5}
\end{equation}

\begin{equation}
y_i \in \big\{ 0 , 1 \big\} \hspace{10 mm}\forall i \in I
\label{c6}
\end{equation}

\begin{algorithm}[t]
\caption{CAGA \Big($p$, $I$, $J$\Big)}
\label{caga}
\renewcommand{\arraystretch}{3.5}
\begin{algorithmic}[1]
\STATE Initialize: $x = 0$, $T = I$, Placed \acp{BBU} $B = \emptyset$, \ac{RRH} Assignment $M_{IJ}: I \longrightarrow J$
\STATE Sort $J$ in increasing values of initial cost $f_j$
\FOR{BBU $j \in J$}
\IF {$(x \leq p)$ $\wedge$ $(T \neq \emptyset)$}
\STATE Sort $T$ in increasing values of front haul cost $c_{ij}$
\FOR{RRH $i \in T$}
\IF{$\delta_i < \eta_j$}
\IF{$B$ DoesNotContain $j$}
\STATE  \textbf{BBU Placement}: Add $j$ to $B$
\STATE $x = x + 1$
\ENDIF
\STATE \textbf{RRH Assignment}: $M_{IJ}: i \longrightarrow j$
\STATE $\eta_j = \eta_j - \delta_i$
\STATE Remove $i$ from $T$
\ENDIF
\ENDFOR
\ENDIF
\ENDFOR
\IF{$T \neq \emptyset$}
\STATE Return: \textbf{Placement and Assignment Failed}
\ELSE
\STATE Return: $B$, $M_{IJ}$.
\ENDIF
\end{algorithmic}
\end{algorithm}

\section{Greedy Approximation Algorithm}\label{greedy}

The \ac{BILP} formulation in \eqref{obj1} - \eqref{c6} could be solved by using one of the available commercial BILP solvers. However, the formulated BILP can be reduced to some known NP-Hard problems. For example, if we simplify the objective function and eliminate some constraints and variables such as \eqref{c2}, \eqref{c4} and \eqref{c5}, the resulting sub-problem can be reduced to the \ac{MCLP} which is known to be NP-hard \cite{MCLP74}. By extension, this implies that the much harder \ac{BILP} in \eqref{obj1} - \eqref{c6} would be computationally intractable for networks of realistic sizes.

In this section, we propose a Cost-Aware Greedy Algorithm (CAGA) that is more computationally viable for larger instances of the problem. CAGA is aimed at minimizing the combined cost of BBU servers and front haul links. CAGA works as follows: we start by sorting the possible BBU servers in ascending values of the installation cost. For each of the BBUs, we also sort the RRH sites according to the cost of creating a front haul link between the BBU and the RRH. Then, choosing the BBUs with the lowest cost, we assign RRHs to it starting with those with the lowest front haul link cost, until the processing capacity of the server is used up. This is repeated until either all the RRHs have been assigned, or the maximum BBU budget $p$ is reached before assigning all RRHs, in which case the algorithm would have failed to find a feasible solution. The psuedo code for CAGA is shown in Algorithm \ref{caga}.\\

\begin{figure*}[ht!]
\setlength{\abovecaptionskip}{7pt plus 0pt minus 0pt}
\setlength{\belowcaptionskip}{7pt plus 0pt minus 0pt}
\begin{minipage}{.33\textwidth}
\centering
\resizebox{.99\textwidth}{!}
{\includegraphics{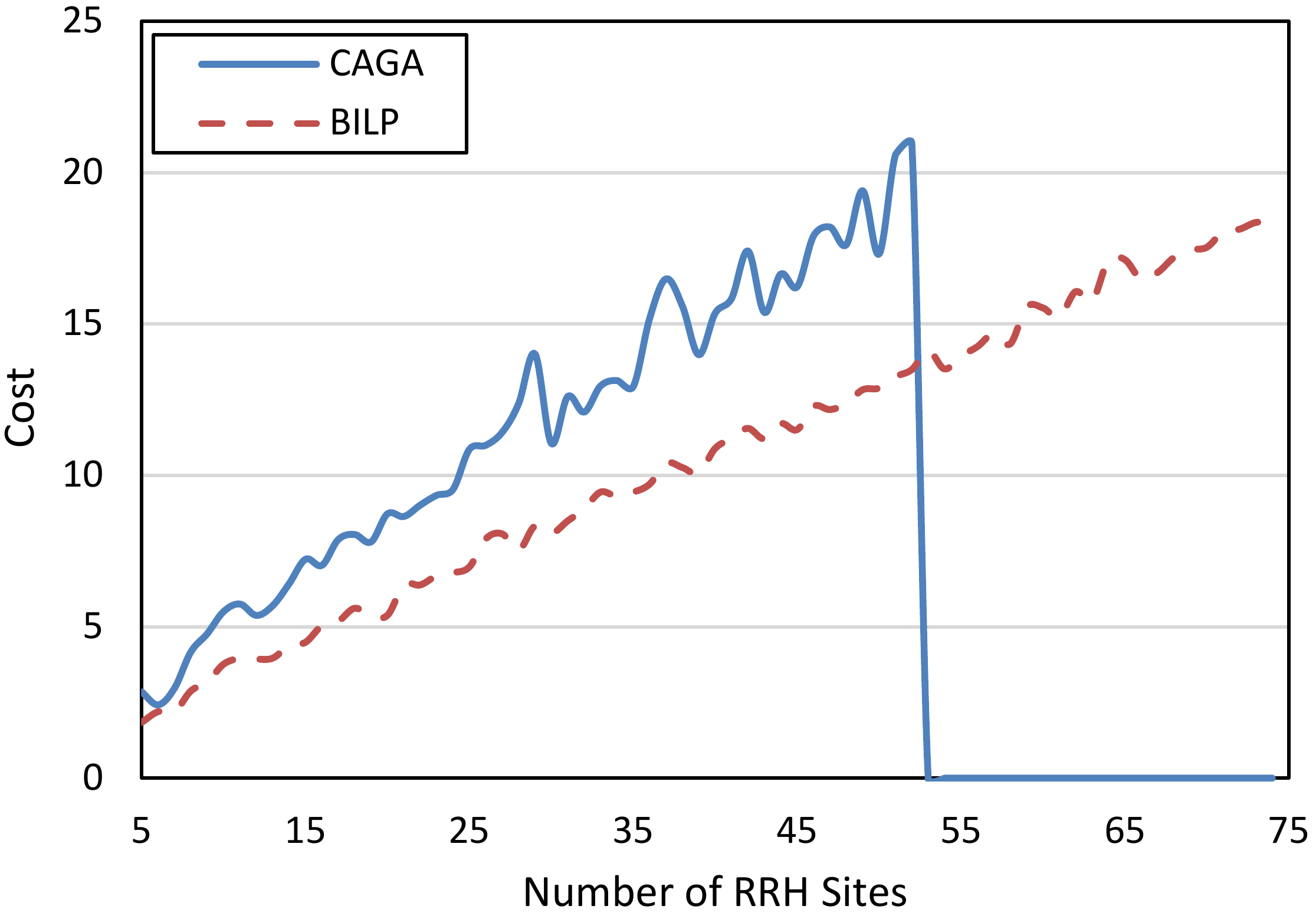}}
  \caption{Placement and Assignment Cost}
  \label{cost}
\end{minipage}
\begin{minipage}{.33\textwidth}
\centering
\resizebox{0.99\textwidth}{!}
{\includegraphics{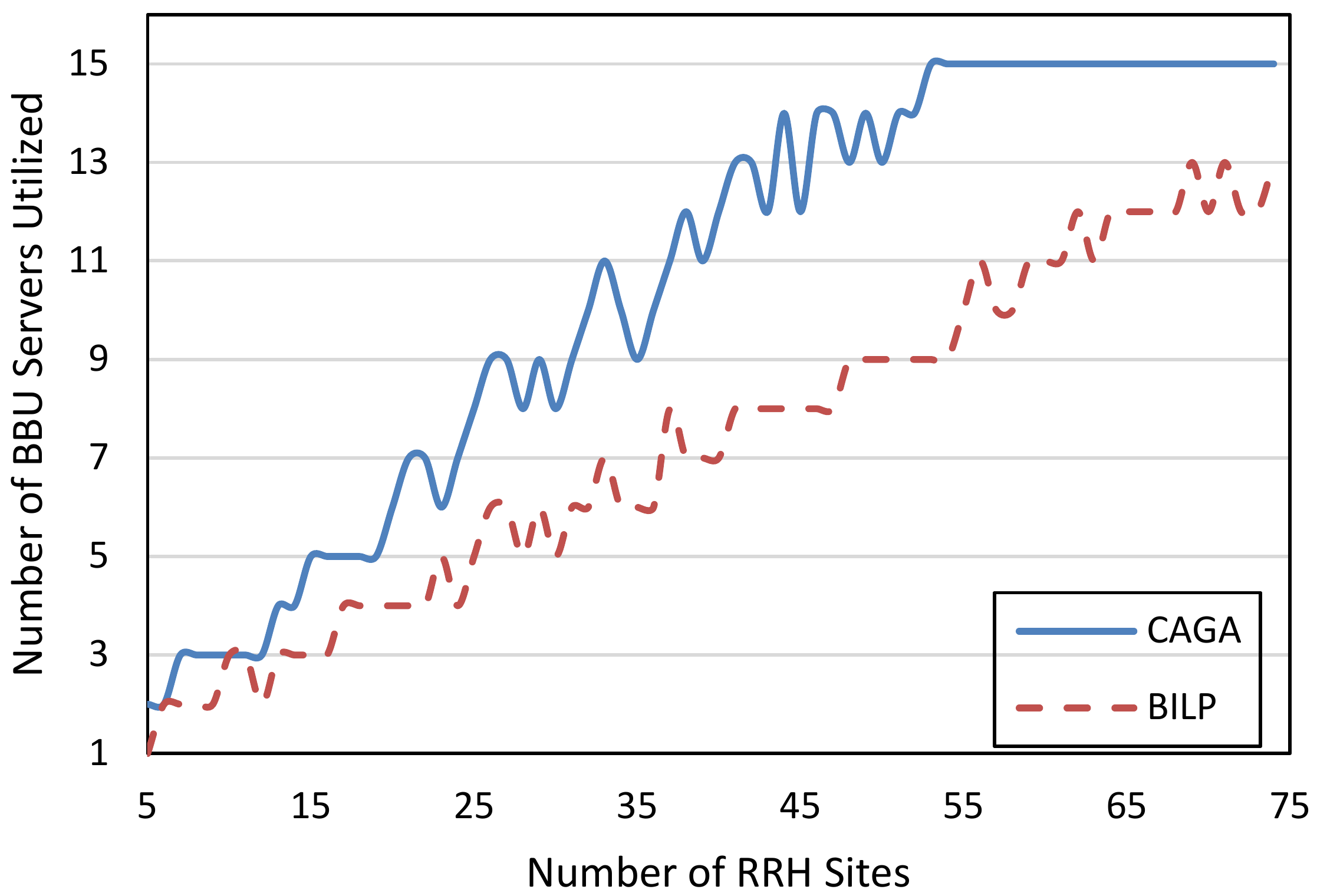}}
  \caption{Actual BBUs Placed}
  \label{numserv}
\end{minipage}
\begin{minipage}{.33\textwidth}
\centering
\resizebox{0.99\textwidth}{!}
{\includegraphics{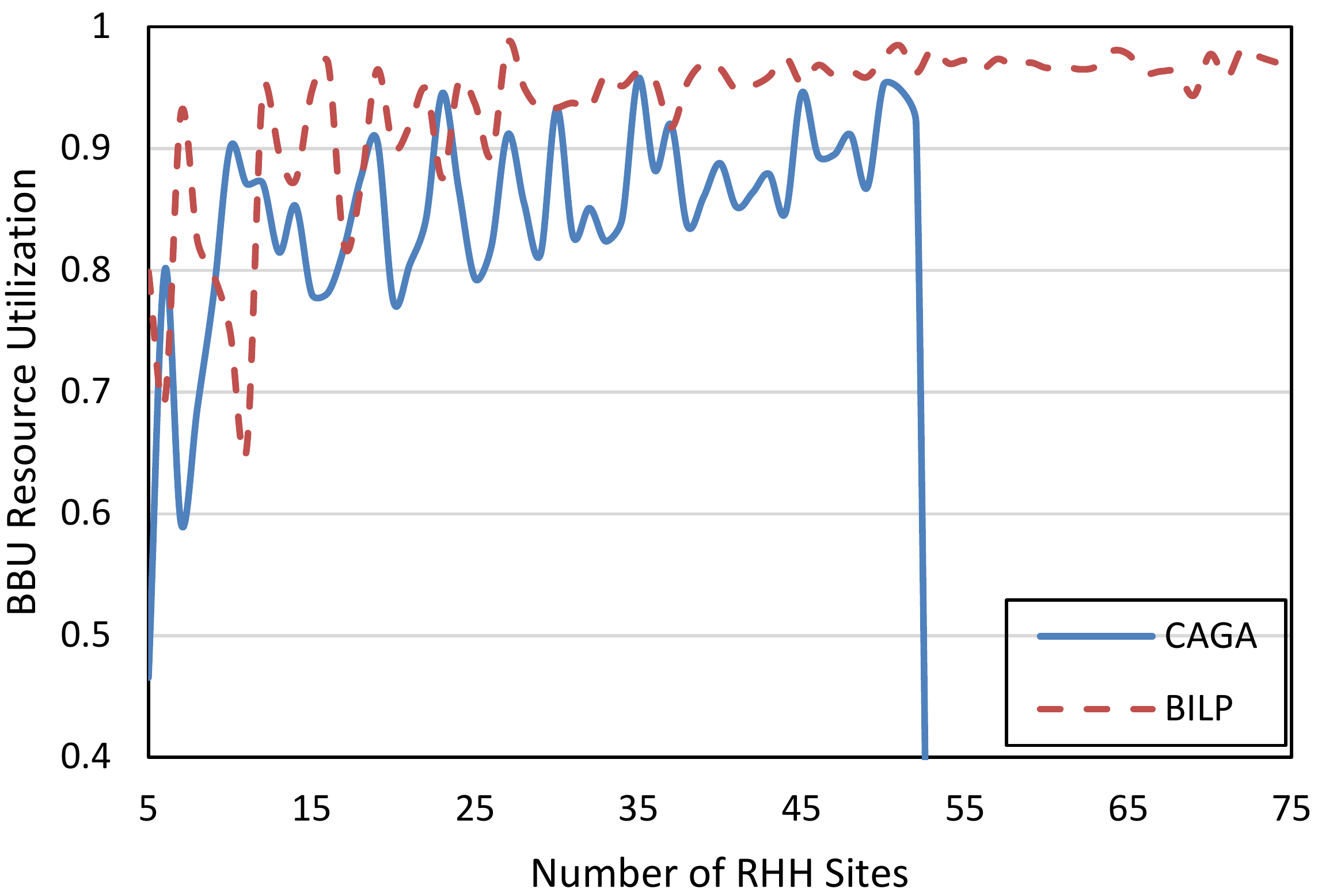}}
  \caption{BBU Average Resource Utilization}
  \label{util}
\end{minipage}\\\\\\

\begin{minipage}{.33\textwidth}
\centering
\resizebox{.99\textwidth}{!}
{\includegraphics{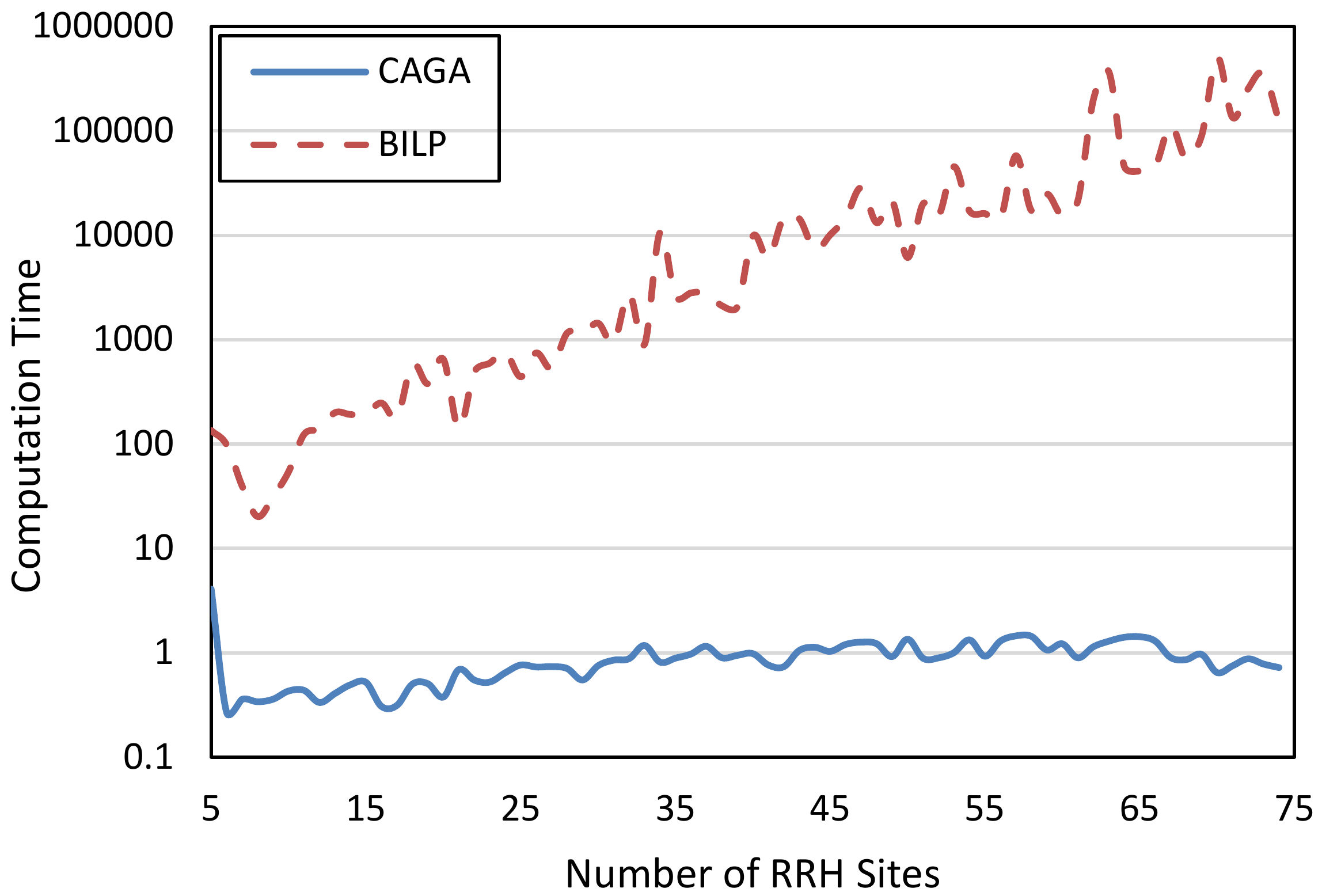}}
  \caption{Computation Time}
  \label{time}
\end{minipage}
\begin{minipage}{.33\textwidth}
\resizebox{.99\textwidth}{!}
{\includegraphics{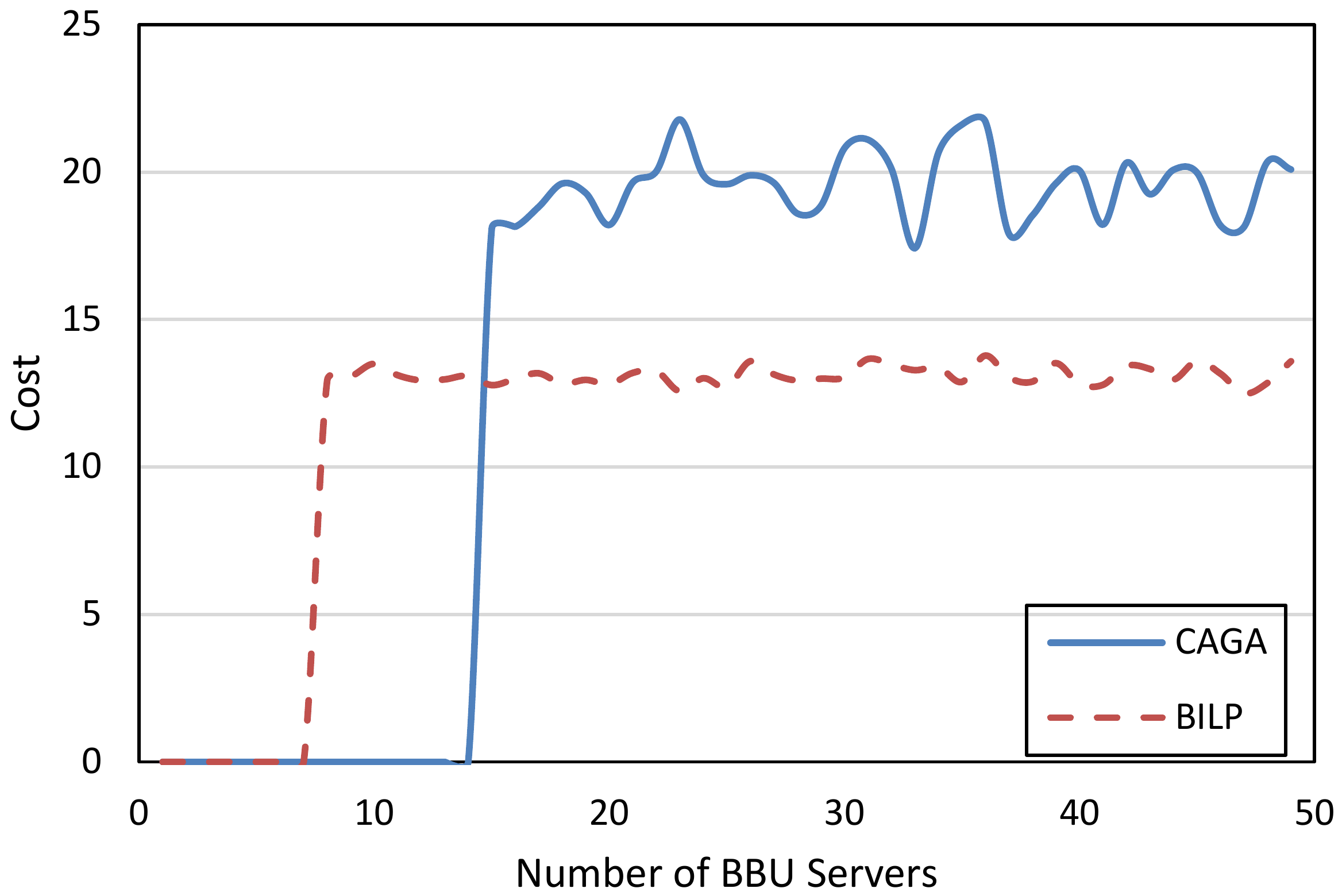}}
  \caption{Effect of Changing Budget}
  \label{changenumserv}
\end{minipage}
\begin{minipage}{.33\textwidth}
\resizebox{.99\textwidth}{!}
{\includegraphics{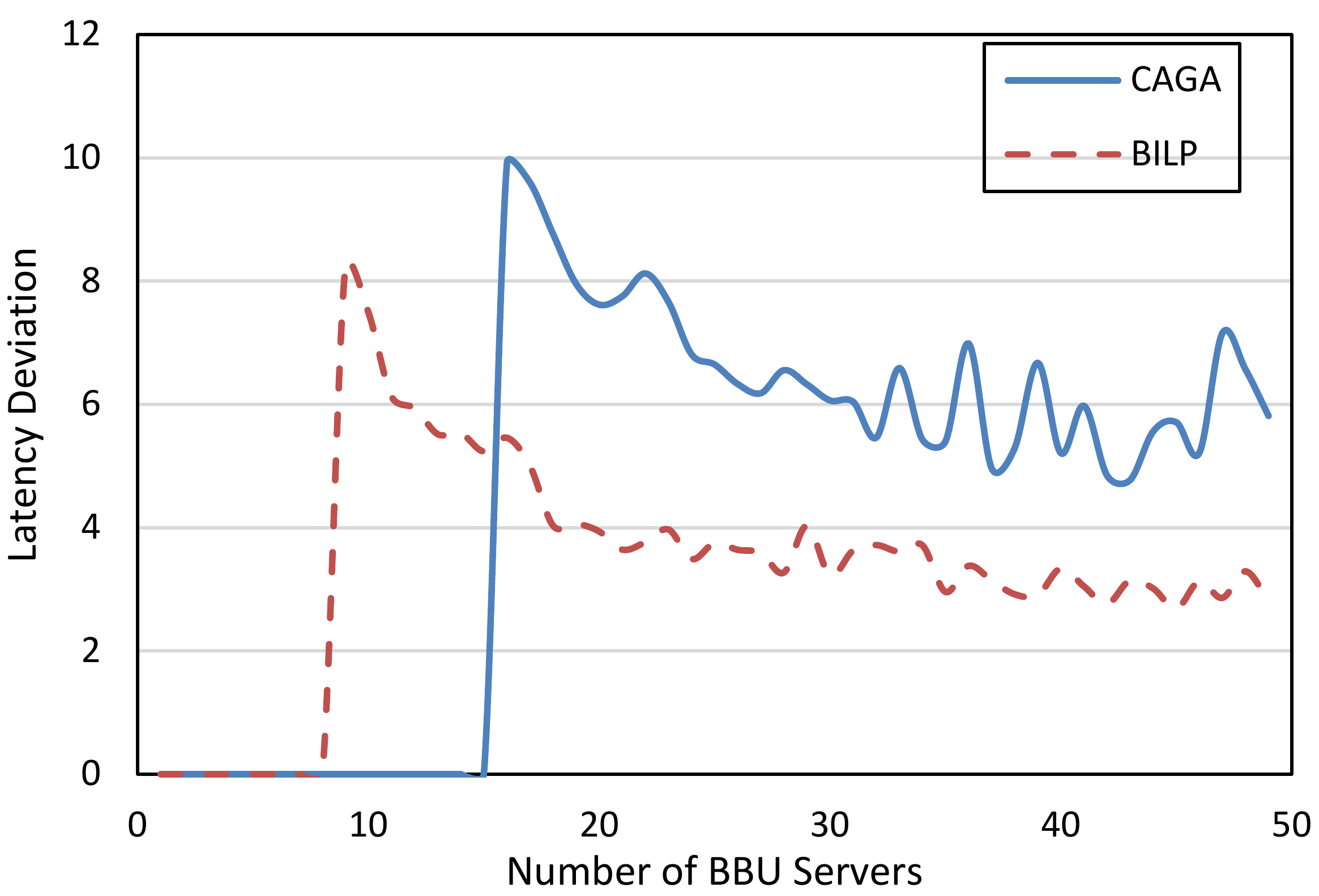}}
  \caption{RHH to BBU Latency}
  \label{latency}
\end{minipage}
\end{figure*}


\noindent \textbf{Time Complexity:} Since the main objective of CAGA is to achieve better time complexity, we now analyze the time complexity of Algorithm \ref{caga}. Line 2 involves sorting of a vector of size $|J|$, which has an average computation complexity of $O(|J| log |J|)$ using the quicksort algorithm \cite{Alsuwaiyel98}. The for loop between lines 6 and 16 involves a maximum of 5 operations (on lines 9, 10, 12, 13 and 14) performed a maximum of $|I|$ times, and can therefore be performed in time $5|I|$. Line 5 also involves sorting of a vector whose maximum possible size is $|I|$, and hence has a maximum average computation complexity of $O(|I| log |I|)$ using the quicksort algorithm. Therefore, the for loop from line 3 to line 18 may have a maximum number of $|J|$ operations, each involving  $O(|I| log |I|)$ plus $5|I|$ operations. This gives a computation complexity of $O\Big(|J| \big(|I| log |I| + 5|I|\big)\Big)$ for the lines 3 - 18, which is the dominating factor in CAGA. This gives an overall polynomial time computation of $O\Big(|I||J|\big(log |I|\big)\Big)$ for the proposed greedy algorithm.

\section{Evaluation} \label{eval}
\subsection{Simulation Setup}
In order to evaluate the proposed algorithms, a simulator was programmed in Java. The RAN was generated on a 500 by 500 grid, with its topology determined using Brite with similar setting as those in \cite{rashidpagevine}. Where no direct link exists between nodes $i$ and $j$, the front haul link, and hence its cost $c_{ij}$ are determined as a sum of costs of creating the front haul link along the shortest path over the RAN topology from $i$ to $j$ using Dijkstra's algorithm. We used the tool ILOG CPLEX 12.60 \cite{CPLEX12.6} to solve the BILP. Simulations were run on Windows 8.1 Pro running on a 16.00GB RAM, Intel i7 4.8GHz Processor Machine. The processing capacities of BBU servers are uniformly distributed between 250 and 500 units respectively, while the processing demand for RRH sites is uniformly distributed between 50 and 100 units. Each RRH site has a desired latency determined from a uniform distribution between $1\times 10^{-7}$ and $1\times ^{-6}$ Units. The fixed cost $u_j$ or $\omega_{ij}$ of each BBU server or each hop of a front haul link respectively is 500 units. Constants $\eta_j$, $\gamma$ and $\chi$ are set to default values of 1.

\subsection{Results}
Fig. \ref{cost} shows the variation of total BBU servers' and front haul links' cost with a changing size of RAN. As the number of RRH sites to be served increases, the costs increase. This is attributed to increases in the number of required front haul links as well as BBU servers to serve them. As expected, it can also be observed that the BILP performs better than CAGA. In fact, beyond a given number of RRH sites, it can be noted that the cost of CAGA reduces to zero. This is explained by the fact that at that point, CAGA is no longer able to find placement and assignment solution as the maximum allowed budget has been reached. For this simulation the maximum budget was 15 BBU servers. Fig. \ref{numserv} confirms this by showing the actual number of utilized BBU servers. It is evident that CAGA reaches the maximum allowed number faster than BILP.\\
\indent In Fig. \ref{util}, we represent resource utilization. This is defined as the proportion of the total processing resources of all placed BBUs which have been assigned to all the RRHs assigned to them. We observe that while both approaches achieve a utilization ratio close to 1, that of BILP is not only slightly higher, but is also more stable. As expected, the resource utilization drops to zero when CAGA is not able to produce feasible solutions. Once again, the different profiles can be attributed to the superior efficiency of BILP which ensures that each placement and assignment utilizes the most appropriate of the available resources.\\
\indent In Fig. \ref{time}, we represent (log scale) the computation times of the two approaches. Compared to BILP, it can be seen that the time required to find a solution using CAGA is not only lower, but also almost un affected by increasing the number of RRH sites. In fact, even when the algorithm is not able to find a feasible solution it utilizes a the same amount of time to determine this. This superiority in time complexity is not surprising since BILP is expected to become computationally intractable for larger problem instances. Therefore, Figs. \ref{cost} and \ref{time} represent the trade-off that has to be taken either for solution quality or for faster solution. The decision on which of the two to utilize may depend both on network size as well as how often the placement and/or assignment has to be done.\\
\indent Finally, Figs. \ref{changenumserv} and \ref{latency} show the effect of varying the budget (maximum allowed number of BBU servers) on both the cost as well as latency. We define latency deviation as the average of all differences between the desired latency of a given RRH site and what is actually obtained after it is assigned to a given BBU. These simulations are performed for a fixed number of 25 RRH sites. We observe that initially, both cost and latency are zero. This is because a very low number of BBU servers does not produce enough resources to serve all the RRH sites. In addition, it can be observed that owing to its superior placement and assignment quality, BILP is able to produce feasible solutions earlier than CAGA. In addition, with regard to the latency, we observe that as the number of available BBUs increases, the latency reduces. This can be explained since when we have more BBU servers, there is more flexibility such that RRUs are assigned to those BBU servers closest to them.

\section{Conclusion} \label{conc}
In this paper, we have formally defined the problem of placing and assigning resources in virtualized radio access networks. We also formulated a binary integer linear programming formulation of the same, and proposed a greedy algorithm for solving bigger instances of the problem. Simulations have shown that these two algorithms allow us to trade solution simplicity and enhanced computation time for better resource management. We hope that these algorithms could be used as a starting point to designing more advanced heuristics that share both the characteristics of solution quality and time complexity.

\section*{Acknowledgment}
This work is partly funded by FLAMINGO, a Network of Excellence project (318488) supported by the European Commission under its Seventh Framework Programme, and project TEC2012-38574-C02-02 from Ministerio de Economia y Competitividad.

\bibliographystyle{IEEEtran}
\bibliography{IEEEabrv,serverlocationbiblio}

\begin{thebibliography}{10}
\providecommand{\url}[1]{#1}
\csname url@samestyle\endcsname
\providecommand{\newblock}{\relax}
\providecommand{\bibinfo}[2]{#2}
\providecommand{\BIBentrySTDinterwordspacing}{\spaceskip=0pt\relax}
\providecommand{\BIBentryALTinterwordstretchfactor}{4}
\providecommand{\BIBentryALTinterwordspacing}{\spaceskip=\fontdimen2\font plus
\BIBentryALTinterwordstretchfactor\fontdimen3\font minus
  \fontdimen4\font\relax}
\providecommand{\BIBforeignlanguage}[2]{{%
\expandafter\ifx\csname l@#1\endcsname\relax
\typeout{** WARNING: IEEEtran.bst: No hyphenation pattern has been}%
\typeout{** loaded for the language `#1'. Using the pattern for}%
\typeout{** the default language instead.}%
\else
\language=\csname l@#1\endcsname
\fi
#2}}
\providecommand{\BIBdecl}{\relax}
\BIBdecl

\bibitem{nfv}
R.~Mijumbi, J.~Serrat, J.-L. Gorricho, N.~Bouten, F.~De~Turck, and R.~Boutaba,
  ``Network function virtualization: State-of-the-art and research
  challenges,'' 2015.

\bibitem{MijumbiNFV15}
R.~Mijumbi, J.~Serrat, J.-L. Gorricho, N.~Bouten, F.~De~Turck, and S.~Davy,
  ``Design and evaluation of algorithms for mapping and scheduling of virtual
  network functions,'' in \emph{IEEE Conference on Network Softwarization
  (NetSoft). University College London}, April 2015.

\bibitem{ChinaMobile}
{China Mobile Research Institute}, ``{C-RAN: The Road Towards Green RAN. White
  Paper. Version 2.5.}'' October 2011.

\bibitem{ETSIUseCases}
{ETSI Industry Specification Group on NFV}, ``{Network Function Virtualization
  (NFV): Use Cases},'' October 2013.

\bibitem{ChihLin14}
C.-L. I, J.~Huang, R.~Duan, C.~Cui, J.~Jiang, and L.~Li, ``Recent progress on
  c-ran centralization and cloudification,'' \emph{Access, IEEE}, vol.~2, pp.
  1030--1039, 2014.

\bibitem{SeokPark13}
S.-H. Park, O.~Simeone, O.~Sahin, and S.~Shamai, ``Robust and efficient
  distributed compression for cloud radio access networks,'' \emph{Vehicular
  Technology, IEEE Transactions on}, vol.~62, no.~2, pp. 692--703, Feb 2013.

\bibitem{LiJPeng14}
J.~Li, M.~Peng, A.~Cheng, Y.~Yu, and C.~Wang, ``Resource allocation
  optimization for delay-sensitive traffic in fronthaul constrained cloud radio
  access networks,'' \emph{Systems Journal, IEEE}, vol.~PP, no.~99, pp. 1--12,
  2014.

\bibitem{MCLP74}
\BIBentryALTinterwordspacing
R.~Church and C.~ReVelle, ``\BIBforeignlanguage{English}{The maximal covering
  location problem},'' \emph{\BIBforeignlanguage{English}{Papers of the
  Regional Science Association}}, vol.~32, no.~1, pp. 101--118, 1974. [Online].
  Available: \url{http://dx.doi.org/10.1007/BF01942293}
\BIBentrySTDinterwordspacing

\bibitem{vygen2005approximation}
J.~Vygen, \emph{Approximation algorithms facility location problems}.\hskip 1em
  plus 0.5em minus 0.4em\relax Forschungsinstitut f{\"u}r Diskrete Mathematik,
  Rheinische Friedrich-Wilhelms-Universit{\"a}t, 2005.

\bibitem{Kenyon97}
C.~Kenyon, ``Best-fit bin-packing with random order,'' in \emph{In 7th Annual
  ACM-SIAM Symposium on Discrete Algorithms}, 1997, pp. 359--364.

\bibitem{VuNguyen14}
V.~N. Ha, L.~B. Le, and N.-D. Dao, ``Cooperative transmission in cloud ran
  considering fronthaul capacity and cloud processing constraints,'' in
  \emph{Wireless Communications and Networking Conference (WCNC), 2014 IEEE},
  April 2014, pp. 1862--1867.

\bibitem{JianHua14}
J.~Tang, W.~P. Tay, and T.~Quek, ``Cross-layer resource allocation in cloud
  radio access network,'' in \emph{Signal and Information Processing
  (GlobalSIP), 2014 IEEE Global Conference on}, Dec 2014, pp. 158--162.

\bibitem{YeguiCia14}
Y.~Cai, F.~Yu, and S.~Bu, ``Cloud radio access networks (c-ran) in mobile cloud
  computing systems,'' in \emph{Computer Communications Workshops (INFOCOM
  WKSHPS), 2014 IEEE Conference on}, April 2014, pp. 369--374.

\bibitem{Yuanming14}
Y.~Shi, J.~Zhang, and K.~Letaief, ``Group sparse beamforming for green
  cloud-ran,'' \emph{Wireless Communications, IEEE Transactions on}, vol.~13,
  no.~5, pp. 2809--2823, May 2014.

\bibitem{Sundaresan135}
\BIBentryALTinterwordspacing
K.~Sundaresan, M.~Y. Arslan, S.~Singh, S.~Rangarajan, and S.~V. Krishnamurthy,
  ``Fluidnet: A flexible cloud-based radio access network for small cells,'' in
  \emph{Proceedings of the 19th Annual International Conference on Mobile
  Computing \&\#38; Networking}, ser. MobiCom '13.\hskip 1em plus 0.5em minus
  0.4em\relax New York, NY, USA: ACM, 2013, pp. 99--110. [Online]. Available:
  \url{http://doi.acm.org/10.1145/2500423.2500435}
\BIBentrySTDinterwordspacing

\bibitem{BastaA2014}
A.~Basta, W.~Kellerer, M.~Hoffmann, H.~J. Morper, and K.~Hoffmann, ``{Applying
  NFV and SDN to LTE Mobile Core Gateways, the Functions Placement Problem},''
  in \emph{{Proceedings of the 4th Workshop on All Things Cellular: Operations,
  Applications, \&\#38; Challenges}}, ser. AllThingsCellular '14.\hskip 1em
  plus 0.5em minus 0.4em\relax New York, NY, USA: ACM, 2014, pp. 33--38.

\bibitem{Moens14}
H.~Moens and F.~D. Turck, ``Vnf-p: A model for efficient placement of
  virtualized network functions,'' in \emph{{Network and Service Management
  (CNSM), 2014 10th International Conference on}}, Nov 2014, pp. 418--423.

\bibitem{Bagaa14}
M.~Bagaa, T.~Taleb, and A.~Ksentini, ``{Service-aware network function
  placement for efficient traffic handling in carrier cloud},'' in
  \emph{{Wireless Communications and Networking Conference (WCNC), 2014 IEEE}},
  April 2014, pp. 2402--2407.

\bibitem{Xia15}
M.~Xia, M.~shirazipour, Y.~Zhang, H.~Green, and A.~Takacs, ``{Network Function
  Placement for NFV Chaining in Packet/Optical Datacenters},'' \emph{{Lightwave
  Technology, Journal of}}, vol.~PP, no.~99, pp. 1--1, 2015.

\bibitem{Yoshida14}
M.~Yoshida, W.~Shen, T.~Kawabata, K.~Minato, and W.~Imajuku, ``{MORSA: A
  multi-objective resource scheduling algorithm for NFV infrastructure},'' in
  \emph{{Network Operations and Management Symposium (APNOMS), 2014 16th
  Asia-Pacific}}, Sept 2014, pp. 1--6.

\bibitem{Rappaport13}
A.~Rappaport and D.~Raz, ``Update aware replica placement,'' in \emph{Network
  and Service Management (CNSM), 2013 9th International Conference on}, Oct
  2013, pp. 92--99.

\bibitem{BariCST13}
M.~Bari, R.~Boutaba, R.~Esteves, L.~Z. Granville, M.~Podlesny, M.~G. Rabbani,
  Q.~Zhang, and M.~F. Zhani, ``Data center network virtualization: A survey,''
  \emph{Communications Surveys Tutorials, IEEE}, vol.~15, no.~2, pp. 909--928,
  2013.

\bibitem{ShidIM15}
R.~Mijumbi, J.~Serrat, and J.-L. Gorricho, ``Self-managed resources in network
  virtualisation environments,'' in \emph{Integrated Network Management (IM),
  2015 IFIP/IEEE International Symposium on}, May 2015, pp. 1099--1106.

\bibitem{path}
R.~Mijumbi, J.~Serrat, J.-L. Gorricho, and R.~Boutaba, ``A path generation
  approach to embedding of virtual networks,'' \emph{Network and Service
  Management, IEEE Transactions on}, vol.~12, no.~3, pp. 334--348, 2015.

\bibitem{rl}
R.~Mijumbi, J.-L. Gorricho, J.~Serrat, M.~Claeys, F.~De~Turck, and
  S.~Latr{\'e}, ``Design and evaluation of learning algorithms for dynamic
  resource management in virtual networks,'' in \emph{Network Operations and
  Management Symposium (NOMS), 2014 IEEE}.\hskip 1em plus 0.5em minus
  0.4em\relax IEEE, 2014, pp. 1--9.

\bibitem{aims}
R.~Mijumbi, J.-L. Gorricho, and J.~Serrat, ``Contributions to efficient
  resource management in virtual networks,'' in \emph{Monitoring and Securing
  Virtualized Networks and Services}.\hskip 1em plus 0.5em minus 0.4em\relax
  Springer Berlin Heidelberg, 2014, pp. 47--51.

\bibitem{sdn}
R.~Mijumbi, J.~Serrat, J.~Rubio-Loyola, N.~Bouten, F.~De~Turck, and
  S.~Latr{\'e}, ``Dynamic resource management in sdn-based virtualized
  networks,'' in \emph{Network and Service Management (CNSM), 2014 10th
  International Conference on}.\hskip 1em plus 0.5em minus 0.4em\relax IEEE,
  2014, pp. 412--417.

\bibitem{neurofuzzy}
R.~Mijumbi, J.-L. Gorricho, J.~Serrat, M.~Shen, K.~Xu, and K.~Yang, ``A
  neuro-fuzzy approach to self-management of virtual network resources,''
  \emph{Expert Systems with Applications}, vol.~42, no.~3, pp. 1376--1390,
  2015.

\bibitem{neural}
R.~Mijumbi, J.-L. Gorricho, J.~Serrat, M.~Claeys, J.~Famaey, and F.~De~Turck,
  ``Neural network-based autonomous allocation of resources in virtual
  networks,'' in \emph{Networks and Communications (EuCNC), 2014 European
  Conference on}.\hskip 1em plus 0.5em minus 0.4em\relax IEEE, 2014, pp. 1--6.

\bibitem{MengXia10}
X.~Meng, V.~Pappas, and L.~Zhang, ``Improving the scalability of data center
  networks with traffic-aware virtual machine placement,'' in \emph{INFOCOM,
  2010 Proceedings IEEE}, March 2010, pp. 1--9.

\bibitem{RabbaniMG13}
M.~Rabbani, R.~Pereira~Esteves, M.~Podlesny, G.~Simon,
  L.~Zambenedetti~Granville, and R.~Boutaba, ``On tackling virtual data center
  embedding problem,'' in \emph{Integrated Network Management (IM 2013), 2013
  IFIP/IEEE International Symposium on}, May 2013, pp. 177--184.

\bibitem{Alsuwaiyel98}
M.~H. Alsuwaiyel, \emph{Algorithms: Design Techniques and Analysis}, ser.
  (Lecture Notes Series on Computing.\hskip 1em plus 0.5em minus 0.4em\relax
  New York, NY, USA: World Scientific Pub Co Inc, 1998, vol.~7.

\bibitem{rashidpagevine}
R.~Mijumbi, J.~Serrat, J.-L. Gorricho, and R.~Boutaba, ``A path generation
  approach to embedding of virtual networks,'' \emph{Network and Service
  Management, IEEE Transactions on}, vol.~PP, no.~99, pp. 1--1, 2015.

\bibitem{CPLEX12.6}
``{IBM ILOG CPLEX Optimizer},''
  \url{http://www-01.ibm.com/software/integration/optimization/cplex-optimizer/about/},
  2015, {Accessed:} 2015-07-19.

\end{thebibliography}

\end{document}